\documentclass[12pt]{article}

\usepackage{graphicx}
\newcommand{\beq}{\begin{eqnarray}}
\newcommand{\eeq}{\end{eqnarray}}

\newcommand{\ket}{\rangle}
\newcommand{\bra}{\langle}

\newcommand{\Thep}{\Theta^+}

\def\lsim{\displaystyle\mathop{<}_{\sim}}

\begin{document}

\begin{center}
{\large  \bf
Decay of $\Thep$ in a quark model}
 
 \vspace*{1cm}
 A. Hosaka\\
 \vspace*{2mm}
 {\it Research Center for Nuclear Physics (RCNP), Osaka University\\
 Ibaraki, Osaka 567-0047, Japan}\\
 \vspace*{0.2cm}
 M. Oka and T. Shinozaki\\
 \vspace*{2mm}
 {\it Department of Physics, Tokyo Institute of Technology\\
 Meguro, Tokyo 152-8551, Japan}\\

\end{center}
\vspace*{0.2cm}

\abstract{
We study the decay of $\Thep$ in a non-relativistic quark model. 
The wave functions are constructed for the two cases
$J^P = 1/2^{\pm}$ as products of 
color, spin, flavor and orbital parts respecting 
total antisymmetrization among the four quarks.  
We find that for the negative parity $\Thep$ the width 
becomes very large which is of order of several hundreds MeV, 
while it is about a several tens MeV for the positive parity.  
By assuming additionally diquark correlations, the width 
is reduced to be of order of 10 MeV or less.  
It is also pointed out that the decay of $3/2^-$ state is
forbidden.  
}

\vspace*{0.5cm}

\vspace*{0.5cm}


\section{Introduction}

One of the distinguished features of the pentaquark particle 
$\Theta^+$ is its very narrow width~\cite{penta}.  
Many experiments so far have been reported only upper 
limit which are less than experimental resolution.  
The pioneering work of the LEPS group at SPring-8 has 
indicated $\Gamma \lsim 25$ MeV~\cite{Nakano:2003qx}, 
while the 
ITEP group has reported $\Gamma \lsim 9$ MeV~\cite{Barmin:2003vv}.  
Recent analysis of the $K^+$ scattering from the xenon or deuteron 
implies even smaller value $\Gamma \lsim 1$ 
MeV~\cite{Cahn:2003wq,Sibirtsev:2004bg,Sibirtsev:2004cf}.  
It has been often mentioned that a width of order of 10 MeV 
or less for baryon resonances is very small 
as compared with a typical value of around 100 MeV, though 
such a criterion should be quantified on a better theoretical 
ground~\cite{Jaffe:2004at}.   
So far the chiral soliton model has predicted the masses and 
widths of the pentaquark baryons
with less theoretical ambiguity based on the SU(3) flavor 
algebra~\cite{Diakonov:1997mm}.  
The model indicates the width of $\Theta^+$ around a few
tens MeV~\cite{Jaffe:2004qj}.  
Then one might wonder if the chiral soliton model does  
something exotic in contrast with the conventional 
knowledge of hadron physics.  

The purpose of this paper is to consider the 
width of $\Theta^+$ in a non-relativistic 
quark model.  
The model has been successful for the description 
of the conventional baryons made dominantly by three quarks.  
The detailed study in the quark model must be useful 
in order to understand the microscopic dynamics of 
the pentaquarks~\cite{Oka:2004xh}.  
Even the result of the chiral soliton model may be 
interpreted just as for the nucleon in the large-$N_c$ 
limit~\cite{Manohar:1984ys}.  
This is, however, beyond our scope in this paper.  
Another question is related to the intrinsic parity of the 
pentaquarks.  
Since we do not know it, we perform the calculation
for the both cases.
As we will see, the decay width depends strongly on the 
parity of $\Thep$.  
Therefore, the study of the decay will help to know the 
parity and hence the internal structure of the pentaquarks.  

In order to prepare for the present study, 
we briefly look at the general aspect for the width of 
baryons in this section.  
Consider a decay of $\Theta^+$ going to the nucleon and 
kaon. 
Assuming the spin of the $\Theta^+$, $J=1/2$, 
the interaction lagrangian takes the form 
\beq
L_{\pm} = g_{KN\Theta} \bar \psi_N \gamma_{\pm} \psi_\Theta K \, , 
\eeq
where $\gamma_{+} = i\gamma_5$ if the parity of $\Theta^+$ is 
positive, while $\gamma_{-} = 1$ if 
the parity of $\Theta^+$ is negative.  
The formula for the decay width is given by 
\beq
\Gamma_+ = \frac{g_{KN\Theta}^2}{2\pi} 
\frac{M_Nq^3}{E_N(E_N+M_N)M_\Theta} \, , 
\label{gamma_p}
\eeq
for the positive parity, where $M_N$ and $M_\Thep$ are the masses 
of the nucleon and $\Theta$, and $E_N= \sqrt{q^2+M_N^2}$ with 
$\vec q$ being the momentum of the final state kaon in the 
kaon-nucleon center of mass system, or equivalently 
in the rest frame of $\Thep$.  
The width for the negative parity is related to the 
one of the positive parity by
\beq
\Gamma_- = \frac{(E_N+M_N)^2}{q^2} \Gamma_+ \, .
\label{gamma_pm}
\eeq
The difference arises due to the different coupling nature:  
p-wave coupling for positive parity $\Theta^+$ and 
s-wave coupling for negative parity $\Theta^+$, representing the 
effect of the centrifugal repulsion in the p-wave.   
In the kinematical point of the $\Theta^+$ decay, 
$M_\Theta = 1540$ MeV, $M_N = 940$ MeV and $m_K = 490$ MeV, 
the factor on the right hand side of (\ref{gamma_pm}) 
becomes about 50 and brings a significant difference 
in the widths of the positive and negative parity
$\Theta^+$.  
If we take $g_{KN\Theta} \sim 10$ 
as a typical strength for strong interaction coupling constants, 
we obtain $\Gamma_+ \sim 100$ MeV, while 
$\Gamma_- \sim$ 5 GeV.  
Both numbers are too large as compared with experimentally 
observed width.  
Therefore, the relevant question is whether some particular 
structure of $\Theta^+$ will suppress the above naive values, or not.  

\begin{figure}
\centerline{\includegraphics[width=12cm]
                        {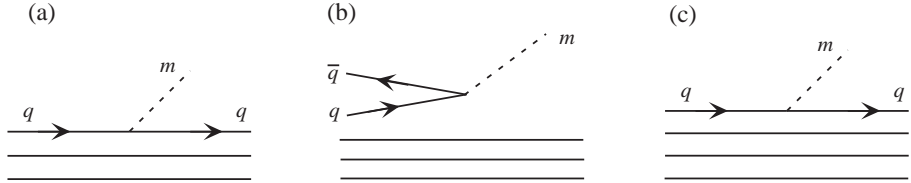}}
\centering
\begin{minipage}{12cm}
\caption{\small 
Meson-baron couplings involving an $mqq$ coupling. 
(a) Transition of a three-quark baryon to another three-quark baryon.
(b) A decay of pentaquark baryon into a three-quark baryon and a meson.
(c) A diagram equivalent to (b).   }
\label{mbb}
\end{minipage}
\end{figure}

In the quark model, assuming that 
the meson, nucleon and pentaquark 
states are dominated by two, three and five valence quarks, 
the decay of the pentaquark 
occurs through the so called fall apart process, in which the 
five quarks dissociate into a three-quark cluster, a nucleon, 
a quark-antiquark cluster, a meson, 
without pair creation of the quarks~\cite{Jennings:2003wz,Close:2004tp}.  
This should be contrasted with an ordinary meson-baryon 
coupling in which a creation of quark-antiquark pair must 
accompany.  
If we treat the meson as a fundamental field, as expected to be
valid for the Nambu-Goldstone boson, 
and introduce a meson-quark interaction of the Yukawa type, 
${\cal L}_{\rm int}$, 
the two processes involve matrix elements of the types
$\bra 0 |{\cal L}_{\rm int} | q\bar q\ket$ for the fall 
apart process, while
$\bra q |{\cal L}_{\rm int} | q\ket$
for the ordinary meson-baryon coupling.  
These are depicted in Figs.~\ref{mbb} (b) and (a), respectively.  
In (a) the coupling is space-like, while in (b) time-like.  
To the extent that the meson is regarded as a point-like, the 
effect of the form factor is neglected.  
This is one of the assumptions we adopt in the present work.  
The quantity we will investigate in this paper is 
essentially the former matrix element 
$\bra 0 |{\cal L}_{\rm int} | q\bar q\ket$
for the decay of $\Thep$~\cite{Diakonov:1997mm}.

In Ref.~\cite{Melikhov:2004qh}, similar calculation was performed, 
where the matrix element of the axial-vector current between 
the $\Theta^+$ and nucleon states were computed.  
In order to relate the matrix element of the axial vector 
current to the axial transition form factor $g_A(\Theta^+ \to N)$, 
they have assumed the PCAC relation, which is, however, 
not applicable to their quark model approach, since the quark model
without the chiral mesons does not generate the meson 
pole term in the current matrix element  
and hence does not satisfy the PCAC relation.

Since we do not know the spin and parity of the $\Thep$, 
we will study the decay width for several spin-parity states.  
Naively,  the negative parity state 
of $(0s)^5$ configuration appears 
lower than positive parity states of $(0s)^4 0p$ configuration.
Although mechanisms which lowers the positive parity states 
have been discussed~\cite{Jennings:2003wz,Stancu:2003if,Hosaka:2003jv}, 
the quantitative prediction for the mass is not yet fully done.  
However, the mass of the $\Thep$ is an important input for 
the decay width, since it changes the phase space volume and also 
the $q$ (momentum) dependent transition form factor.  
In the present study, we use as inputs of masses experimental 
values in order to exclude the dependence coming from the 
phase space and the $q$ dependent form factor.  
In this way, comparison of the results of different states
reflects the difference in the structure of, in particular,  
the internal spin-flavor-color wave functions.  

This paper is organized as follows.  
In section 2, we establish the necessary ingredients 
of the quark model, especially for the basic 
meson-quark interaction and how to compute 
the matrix element of the fall apart process.  
In sections 3, 
we calculate the transition matrix element from the 
five-quark state of $\Thep$ to $KN$ in the non-relativistic
quark model of harmonic oscillator.  
An advantage of the model is that the separation of the 
center of mass coordinate is completely performed.  
We compute the matrix elements for both positive and negative 
parity $\Thep$.  
In the fall apart process 
the decay of $\Thep$ into $KN$ proceeds first by forming 
a nucleon-like $qqq$ state and a kaon-like $q \bar s$ 
state in the $\Thep$ wave function.  
It is then necessary to compute the spectroscopic factor
for a given quark model wave function for $\Thep$. 
This is shown in detail in appendix.  
In section 4 numerical values are presented with some  
discussions.   
In the final section the paper is concluded.  

\section{Ingredients of the quark model}

Our starting point is an interaction lagrangian of chiral 
mesons and quarks~\cite{Manohar:1983md}: 
\beq
{\cal L}_{\rm int} 
=
-i\ g \bar \psi \gamma_5 \Phi \psi 
\sim 
\frac{g}{2m} \chi^\dagger \vec \sigma \cdot \vec \nabla \Phi \chi
\, , 
\label{Lmqq}
\eeq
where $\psi = (\psi_u, \psi_d, \psi_s)$ is 
a four component Dirac spinor field, 
$\chi =(\chi_u, \chi_d, \chi_s)$ the two component spinor field 
and 
\beq
\Phi = \left(
\begin{array}{ccc}
\pi^0 + \frac{1}{\sqrt{3}} \eta &
  \sqrt{2}\pi^+ & \sqrt{2}K^+ \\
\sqrt{2}\pi^- &
  -\pi^0 + \frac{1}{\sqrt{3}} \eta & \sqrt{2}K^0 \\
\sqrt{2}K^- &
  \sqrt{2}\bar K^0 & -\frac{2}{\sqrt{3}} \eta	
\end{array}
\right) \, 
\label{defPhi}
\eeq
is a flavor octet meson field.  
In the second equation of (\ref{Lmqq}), 
we have shown an expression familiar in the 
nonrelativistic quark model with a quark mass $m$. 
The meson-quark coupling constant $g$ may be determined from the 
$\pi NN$ coupling constant $g_{\pi NN} \sim 13$.  
Taking the matrix element of (\ref{Lmqq}) in the nucleon states
of the quark model, we find
\beq
\bra N(p_2) \pi^a | {\cal L}_{\rm int} | N(p_1)\ket 
\sim 
i \frac{5g}{6m} \vec \sigma_N \cdot \vec q \tau^a \, , 
\label{NLN}
\eeq
where $\vec q = \vec p_2 - \vec p_1$ and $\vec \sigma_N$ 
is the spin matrix acting on the two component nucleon 
spinor.  
Assuming that quark mass is 1/3 of nucleon mass, 
$m \sim M_N/3$, and comparing (\ref{NLN}) with the 
$\pi NN$ interaction 
$(i g_{\pi NN}/2M_N) \vec \sigma_N \cdot \vec q$, 
we find 
\beq
g = \frac{g_{\pi NN}}{5} = 2.6 \, .  
\label{gvalue}
\eeq

This interaction has been often used in the quark model to compute 
meson-baryon couplings and transition amplitudes of, for instance, 
$N^* \to \pi N$ where $N^*$ is an ordinary nucleon resonance made 
from three quarks as shown in Fig.~\ref{mbb}~(a).  
The same interaction can be used for the decay of a pentaquark baryon 
if one reverse the outgoing quark line for the incoming antiquark line
as shown in Fig.~\ref{mbb}~(b). 
Here the quark-antiquark pair has the quantum numbers of the kaon.  
When we treat the quarks as identical particles, it is convenient to 
consider the diagram (c), where the antiquark line in the initial 
pentaquark state is once again  
reversed to an outgoing quark line.  
This is the ``particle-hole transformation" which relates the 
interaction between the two particle states with the one for the
particle and hole states~\cite{bohr_mottelson}.

Now the pentaquark $\Thep$ wave function 
$|\Thep\ket$ can be written by four light quarks $uudd$ 
and an $\bar s$.  
The state contains a component of the first 
three quarks having the neutron quantum numbers and the 
remaining quark and antiquark having kaon quantum numbers,
\beq
|\Thep\ket = a |(u(1)d(2)d(3))^n (u(4)\bar s(5))^{K^+}\ket 
+ \cdots \, , 
\eeq
where $a$ is the spectroscopic factor, 
probability amplitude of finding the 
$K^+ n$ state in $|\Theta^+\ket$.  
Then transition amplitude for 
$\Thep \to K^+ n$ can be written as 
\beq
\bra f | \int d^4 x \, {\cal L}_{\rm int} |i\ket 
&=& 2 \pi \delta(E_f - E_i) {\cal M}_{fi}\, , 
\nonumber \\
{\cal M}_{\Theta^+ \to K^+ n} &=& -i \, \bra n_f K^+(\vec q) | 
\int d^3 x \, g \bar \psi \gamma_5 \Phi \psi 
| \Thep(uudd\bar s)\ket \nonumber \\
&=&
-i \sqrt{2} 
\bra n_f(udd) |  \int d^3 x \, g \bar \psi \gamma_5 \psi
e^{-i \vec q \cdot \vec x} 
|\Thep(uudd\bar s)\ket ,
\label{calM1}
\eeq
where $n_f$ denotes the final state neutron of three quarks.  
In the valence quark model, 
In the last line of (\ref{calM1}), 
we have assumed that the final state kaon is expressed by 
a non-interacting plain wave of momentum $\vec q$.  
In practical calculations, we treat the quarks as identical 
particles.  
By moving $\bar s$ in the initial state into $s$ in 
the final state, the initial and final states may be treated as 
systems of four identical particles.  
Then the operator is written as a sum over the four particles
${\cal O} = \sum_{i=1,...,4} {\cal O}(i)$, 
and the final state may be antisymmetrized as 
\beq
| udds\ket = \frac{1}{2}
\left[ n(123) s(4) -  n(124) s(3) - n(143) s(2) - n(423) s(1)
\right]\, ,   
\label{nKfinal}
\eeq
where we have assumed that $n(ijk)$ is already antisymmetrized.  
The four quarks $uudd$ in the initial state $\Thep$ are 
also antisymmetrized having the same structure as (\ref{nKfinal})
under permutation.  
The $\Thep$ wave function is explicitly constructed in the next section.  
Combining (\ref{calM1}) and (\ref{nKfinal}) with correct 
counting factors, we find
\beq
{\cal M}_{\Theta^+ \to K^+ n} &=& 2 \sqrt{2} g a 
\bra 0| 
\int d^3 x \bar \psi \gamma_5 \psi e^{-i\vec q \cdot \vec x}
| (u\bar s)^{K^+}\ket \bra n_f | (udd)^n \ket .  
\label{calM2}
\eeq

We will compute the matrix element of (\ref{calM2}) for both positive 
and negative parity $\Thep$.   
Carlson et al~\cite{Carlson:2003xb} 
computed the constant $a$ for several configurations.  
Here we repeat the calculations briefly using the method 
of Young diagram.

\section{Decay amplitude}

\subsection{Negative parity}

\begin{figure}
\centerline{\includegraphics[width=12cm]
                        {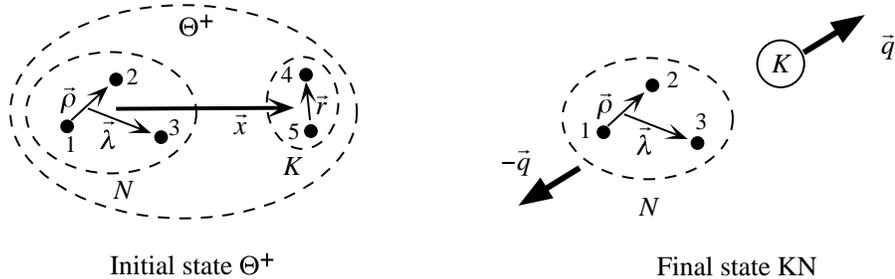}}
\centering
\begin{minipage}{12cm}
\caption{\small 
Definition of various coordinates of the five quark system.  }
\label{fiveq}
\end{minipage}
\end{figure}

Now we compute the matrix element (\ref{calM2}) using a 
wave function of harmonic oscillator for 
the non-relativistic quark model~\cite{hosaka_ws}.  
For simplicity, we assume that 
all quarks are in a potential with the common 
oscillator parameter, 
$m\omega \equiv \alpha_0$.  
Hence the $(0s)^5$ configuration 
for the negative parity state 
may be written as
\beq
|\Theta^+, (0s)^5 \ket &=& 
\psi(\vec x_1) \psi(\vec x_2) \psi(\vec x_3) 
\psi(\vec x_4) \psi(\vec x_5) |\Theta^+_{csf}\ket\, , 
\label{psi5}
\eeq
where the color-spin-flavor wave function is given by 
(\ref{csf_negaiive}) in 
appendix with the spectroscopic factor 
$a = 1/(2\sqrt{2})$.  
The single particle wave function
is given by   
\beq
\psi(\vec x_i) = \left(\frac{\alpha_0^2}{\pi} \right)^{3/4}
\exp\left( - \frac{\alpha_0^2}{2}|\vec x_i|^2 \right) \, .
\eeq
By introducing various coordinates as defined 
in Fig.~\ref{fiveq}, we can decompose the single 
particle state (\ref{psi5}) into a product of 
parts of the corresponding coordinates.  
After the separation of the wave function for the 
total center of mass coordinate $\vec X_{\rm tot}$, and then
replacing it by the plane wave of the total momentum 
$\vec P_{\rm tot} = 0$, we can write 
\beq
|\Theta^+\ket 
&\sim& 
a \, |(u(1)d(2)d(3))^n (u(4)\bar s(5))^{K^+}\ket + \cdots 
\nonumber \\
&=&
a \, e^{i \vec P_{\rm tot} \cdot \vec X_{\rm tot}}
\phi_{KN}(\vec x) \phi_N(\vec \rho, \vec \lambda)
\phi_K(\vec r) 
\times  ({\rm color}) \cdot ({\rm spin}) \cdot ({\rm flavor}) 
+ \cdots \, ,
\eeq
where the color-spin-flavor wave function is presented in appendix, 
and the dots in the last line contains all possible states 
composed of products of color singlet $3q$ and $q \bar q$ 
states which are orthogonal to the $K^+ n$ state in the first term.  
The wave function
$\phi_{KN}(\vec x)$ is for the relative motion of the nucleon 
and kaon like clusters, 
$\phi_N(\vec \rho, \vec \lambda)$ for the intrinsic state of the 
nucleon like part and 
$\phi_K(\vec r)$ for the intrinsic (relative) 
state of the kaon like part.  
For instance, 
\beq
\phi_{KN}(\vec x) = 
\left( \frac{\alpha^2}{\pi}\right)^{3/4}
\exp\left(-\frac{\alpha^2}{2}x^2\right)\, , 
\label{phiKN}
\eeq
where the parameter $\alpha$ is for the 
relative motion of the kaon and nucleon like clusters
and is related to $\alpha_0$ by 
\beq
\alpha^2 = \frac{6}{5} \alpha_0^2\, .
\eeq

The final state wave function takes the form
\beq
e^{i \vec P_{\rm tot} \cdot \vec X_{\rm tot}}
e^{i \vec q \cdot \vec x}
\phi_N(\vec \rho, \vec \lambda)\, , 
\eeq
where we have assumed that the intrinsic structure of the 
final state nucleon is the same as the one of the three quark 
cluster in the initial state, and the
relative motion of the kaon and the nucleon is described by 
a plane wave of momentum $\vec q$.  

The remaining computation is rather straightforward but 
it is worth mentioning a few remarks.  
First, the overlap of the three-quark wave functions in the 
initial and final states is set one by assuming that 
the oscillator parameter of the five quarks in the 
initial state $\Theta^+$ is the same 
as that of the nucleon in the final state. 
Therefore, we have 
$\bra n_f | (udd)^n \ket = 1$.  
If spatial structure for the nucleon and pentaquark states 
are different, this overlap factor will be suppressed from 1.  
Furthermore, a small repulsive force in the $KN$ scattering 
channel also reduce the overlap.   
As discussed in Ref.~\cite{Melikhov:2004qh}, strong diquark
correlation has a significant effect on the suppression.  
Whether sufficient spatial correlation will be developed or not 
is, however, a dynamical question~\cite{Dudek:2004cf}.  
In any case, we expect some suppression in a more realistic study, 
and therefore the estimation 
in the present work provide the upper bound in the quark model 
calculation.  
Second, in the matrix element for the 
annihilation of the kaon like cluster in the initial state
reduces to 
\beq
\bra 0| \bar \psi_{\bar s} \gamma_5 \psi_u
| (u\bar s)^{K^+}\ket 
\sim
\sqrt{2} \left(\frac{\alpha_0^2}{2\pi}\right)^{3/4} 
\phi_K(\vec x)\, .  
\eeq
Here, $\gamma_5$ is replaced by one in the non-relativistic 
approximation, since the lower component of the 
antiquark wave function is a large component.  
The factor $\sqrt{2}$ is from the spin part of 
the matrix element for $S = 0$ pair of $u$ and $\bar s$ quarks whose 
wave function is given by 
$1/\sqrt{2}(\uparrow \downarrow - \downarrow \uparrow)$, and 
$(\alpha_0^2/2\pi)^{3/4}$ is the value of the 
kaon like wave function $\phi_K(\vec r)$ at the origin.  

The resulting the matrix element of (\ref{calM2}) is 
compared with the matrix 
element of the s-wave coupling, 
\beq
\bra n(-\vec q) | g_{KN\Theta} 
\int d^3x \bar \psi_{n} \psi_\Theta e^{-i\vec q \cdot \vec x} 
|\Theta^+(\vec 0) \ket
=
(2\pi)^3 \delta^3(0) g_{KN\Theta}\, , 
\eeq
where we have set the normalization factor 
$\sqrt{(E+M)/2M} \to 1$.  
We find the result 
\beq
g_{KN\Theta} = 4 g a
\left( \frac{5}{3}\right)^{3/4} F(q)\, , 
\label{gKNT_neg}
\eeq
where the transition form factor is defined by 
\beq
F(q) = \left(\frac{\alpha^2}{2\pi}\right)^{3/2} \int d^3 x 
e^{-\frac{1}{2} \alpha^2 \vec x^2} e^{-i\vec q \cdot \vec x}
=
e^{-\frac{q^2}{2\alpha^2}} \, .  
\label{defF}
\eeq 

\subsection{Positive parity}
 
For a positive parity $\Thep$, one of the four light 
quarks must be excited into a p-orbit, and 
hence the five quark 
configuration is $(0s)^4 0p$.  
For this case, four independent 
configuration are available~\cite{Jennings:2003wz}.  
In general, 
the lowest energy configuration may be a linear 
combination of these states.  
Here we consider three simple cases; 
the one minimizing the spin-flavor 
interaction~\cite{Carlson:2003xb}, 
the one minimizing the spin-color interaction
and the one with strong $S=I=0$ diquark 
correlation~\cite{Jaffe:2003sg}.  
For illustration, however, we show detailed computation 
only for the first one of minimizing the spin-flavor 
interaction.  

In the non-relativistic quark model of harmonic oscillator, 
we can write the two terms of the p-states of (\ref{theta_p1})
in terms of $KN$ relative coordinate, as
\beq
|\Theta^+, (0s)^4 0p\ket &=& 
\sqrt{\frac{5}{96}} \alpha \, [\vec x, \chi]_{J=1/2, m_i} 
\left( \frac{\alpha}{2\pi} \right)^{3/2}
e^{-\frac{1}{2} \alpha^2 x^2}
\left(\frac{5}{3}\right)^{3/4}
\nonumber \\
&\times & 
e^{i\vec P_{\rm tot} \cdot \vec X_{\rm tot}} 
\phi_N(\vec \rho, \vec \lambda)\, , 
\label{theta_p2}
\eeq
where we have set $\vec x_4 \to \vec x_5$ prior to 
computation of the matrix element.
In (\ref{theta_p2}), $[\vec x, \chi]_{J=1/2, m_i}$ represents
the coupling of the relative coordinate between the kaon and 
nucleon like clusters $\vec x$ and the spin 
1/2 state $\chi$ to form the total spin $J = 1/2, m_i$.  
Furthermore, we have recovered the spectroscopic factor 
$a = \sqrt{5/96}$ as derived by Carlson et al~\cite{Carlson:2003xb}.  

The matrix element (\ref{calM2}) 
can now be computed with the result
\beq
{\cal M} &=&
(2\pi)^3 \delta^3(0) \sqrt{\frac{5}{96}}
\frac{4g}{\sqrt{3}\alpha} \bra m_f | \vec \sigma \cdot \vec q | m_i\ket 
F(q)
\left( \frac{\alpha}{2\pi} \right)^{3/2}
\left(\frac{5}{3}\right)^{3/4}\, .
\eeq
which is compared with the p-wave coupling defined by 
\beq
g_{KN\Theta} \bra n(-\vec q) | \int d^3 x \, 
\bar \psi_n \gamma_5 \psi_\Theta e^{-i\vec q \cdot \vec x}
|\Theta(\vec 0)\ket = (2\pi)^3 \delta^3(0) g_{KN\Theta} 
\frac{\vec \sigma \cdot \vec q}{2M_N}.  
\eeq
Hence we find 
\beq
g_{KN\Theta} = g \sqrt{\frac{5}{96}}
\frac{8M_N}{\sqrt{3}\alpha} 
\left(\frac{5}{3}\right)^{3/4} F(q)\, ,
\label{g_pos}
\eeq
where $\alpha = \sqrt{6/5}\alpha_0$ and 
$F(q)$ has been defined by (\ref{defF}).  

The appearance of the oscillator parameter 
$\alpha$ in the denominator is worth pointed out.  
It indicates that as the size (inversely proportional 
to $\alpha$) of $\Thep$ decreases, the coupling constant
and hence the decay width decreases.  
This is a feature of the fall apart decay into a
p or higher partial wave state.
For a small 
$\Thep$ the decay is suppressed, since the overlap with the 
decaying p-wave final state is suppressed.
This feature is very much different from the 
decay (transition) of an ordinary baryon which is 
accompanied by the creation of quark-antiquark 
pair for a meson.
Such a decay remains finite in the limit that the size 
of the baryon is zero.  

\section{Numerical values and discussions}

The decay width is given by the square of the coupling constant
times the phase space volume.   
Since the change in the mass affects the phase
space volume and the transition form factor,   
our study here is considered to be for the 
coupling constant $g_{KN\Theta}$ at the realistic 
kinematical point by fixing the 
masses at experimental values:  
$m_K \sim 490$ MeV, $M_N = 940$ MeV and $M_\Theta = 1540$ MeV. 
For numerical estimation, we consider the following 
two cases corresponding to different sizes of harmonic 
oscillator potential:
\beq
\bra r^2 \ket_N^{1/2} = 1.0\; {\rm fm} &\to&
\alpha_0^2 = 1.5 \; {\rm fm}^{-2} \, , \nonumber \\
\bra r^2 \ket_N^{1/2} = 0.7\; {\rm fm} &\to&
\alpha_0^2 = 3.0 \; {\rm fm}^{-2} \, .
\eeq
The resulting coupling constants and the decay widths 
are summarized in Table 1.  
As discussed before, since we do not consider possible 
difference in the structures of the spatial wave functions
of the nucleons and pentaquark, the values in Table 1 should 
be upper bounds.

From the table,  we see that the width of the negative 
parity $\Thep$ is too wide for the state to be regarded as 
a sharp resonance, as consistent with the naive estimate 
made in section 1.    
In this paper, we have shown this by explicitly calculating 
the matrix element.  
However, for the ground state configuration $(0s)^5$, 
this could have been expected, if we have noticed 
that this is the unique configuration, unlike the 
positive parity $\Thep$.  
Due to this uniqueness, the $\Thep$ wave function 
is completely written as a $KN$ like state as given 
in (\ref{nKfinal}).  
The spectroscopic factor $1/2$ (for finding two $KN$
states) is then identical to the normalization factor 
of (\ref{nKfinal}).  
Hence, unless there is some attraction and/or   
coupled channels, $1/2^-$ state can not accommodate 
a resonance~\cite{Jaffe:2004at,Jennings:2003wz}.  
If one could include a higher excited state such as 
a $(0s)^4 1s$ configuration, there could be a resonance
state, but the energy would become very large.  

For the positive parity $\Thep$, the column SF (spin-flavor) 
shows the results for the $\Thep$ configuration minimizing 
the spin-flavor interaction as we have discussed so far.  
The column SC (spin-color) is for the result for the 
configuration minimizing the spin-color interaction.  
The SC configuration has a spectroscopic factor 
$\sqrt{5/192}$
which is smaller than that of the SF by the factor 
$1/\sqrt{2}$. 
Therefore, the expected decay width becomes half.   
We have also shown in the column JW the result for the 
case where the Jaffe-Wilzeck type of diquark correlation 
is developed~\cite{Jaffe:2003sg}.  
In this case, the spectroscopic factor becomes 
$\sqrt{5/576}$~\cite{Carlson:2003xb}
instead of $\sqrt{5/96}$, which reduces the decay width 
by the factor 6 from the result of SF.  

Although the results depend significantly on 
the choice of model parameters, 
the general tendency is that the decay width for the 
negative parity $\Theta^+$ is too wide, 
while that for the positive parity can be of order of 
10 MeV when strong correlation in the color-spin-flavor space 
is developed. 
As anticipated spatial correlation will 
further suppresses the decay width.  
Interference between different configurations could be 
another source of reduction.
This, however, is a difficult problem at this point, 
since it depends very much on the type of interaction.  

The present analyses can be extended straightforwardly 
to the case of spin 3/2.  
For the negative parity case, the spin 1 state of the four quarks
in the $\Thep$ may be combined with the spin of $\bar s$ for 
the total spin 3/2.  
In this case the final $KN$ state must be in d-wave, and therefore, 
the spectroscopic factor of finding a d-wave $KN$ state in the 
initial configuration which is $(0s)^5$ is simply zero.  
If a tensor interaction induces a small admixture of 
a d-wave configuration, it can decay into a d-wave $KN$ state.  
The decay rate, however, would be small.  
Therefore the decay of $3/2^-$ $\Thep$ into $KN$ is expected to be 
suppressed.  
There could be a possible decay channel of the nucleon and the 
vector $K^*$ of $J^P = 1^-$.  
This decay, however, does not occur 
since the total mass of the decay channel is larger than 
the mass of $\Thep$.  
Hence this $J^P = 3/2^-$ state could be another 
candidate for the observed narrow state.  
This state does not have a spin-orbit partner and forms 
a single resonance peak around its energy.  

For the positive parity case, the p-state orbital excitation 
may be combined with the spin of $\bar s$ for the total spin 3/2. 
In this case, the calculation of the decay width is precisely 
the same as before except for the last step of 
Eqs. (\ref{c_of_p}) and (\ref{theta_p1}), where the total 
spin should be 3/2.  
After taking the average over the angle $\vec q$, 
however, the coupling 
yields the same factor as for the case $J = 1/2$.  
Hence the decay rate of spin 3/2 $\Thep$ is the same as
that of $\Thep$ of spin 1/2 in the present simple treatment.  

\begin{table}[tbp]
\centering
\caption{\label{widthp} \small The $KN\Thep$ coupling 
constant $g_{KN\Thep}$ and decay width (in MeV) 
of $\Thep$ for $J^P = 1/2^{\pm}$.  
In the columns of SF, SC and JW presented are the results for the 
configuration that minimize the spin-flavor interaction, 
that minimizes the spin-color interaction and that with the 
$S = I = 0$ diquark correlation of the Jaffe-Wilczek type. }
\vspace*{0.5cm}
{\small 
\begin{tabular}{ c c | c c c c | c c c c}
\hline
 &  & \multicolumn{4}{c |}{$g_{KN\Theta}$} & 
          \multicolumn{4}{c}{$\Gamma$ (MeV)} \\
 &  & $J^P=1/2^-$ & & $1/2^+$ & & $J^P=1/2^-$ & 
 & $1/2^+$ & \\
\hline
$\bra r^2 \ket^{1/2}$ &  $\alpha_0^2$  &   & SF   &   SC   &   JW
          &   & SF   &   SC   &   JW   \\
$1/\sqrt{2}$ fm & 3 fm$^{-2}$ & 4.1 & 7.7 & 5.5 & 3.2 & 
  890 & 63 & 32 & 11 \\
1 fm & 1.5 fm$^{-2}$ & 3.2 & 8.4 & 5.9 & 3.4& 520 & 74 & 37 & 12 \\
\hline
\end{tabular}
}
\end{table}

\section{Conclusion}

We have studied the decay of the pentaquark baryon $\Thep$
in the non-relativistic quark model.
The matrix element of the fall apart process has been 
calculated using the meson-quark coupling of the Yukawa type.  
The method is a natural extension of the standard 
quark model calculation for meson-baryon couplings which 
involves the matrix element of the axial current 
between quark states to the one of quark-antiquark 
annihilation.  
If the quark-meson coupling would be fundamental, 
such an approach should be reliable.  

In the quark model, we can consider 
various states, in the present study a positive and negative 
parity pentaquark states.  
This is perhaps an advantage over the chiral soliton model, 
since in the latter negative parity states require an excitation 
beyond the rigid rotation, which is rather difficult to construct.    
In the quark model of harmonic oscillator,  
ground state of $(0s)^5$ is dominated by the 
$KN$ scattering state and is necessarily lead to a 
large decay width of the $1/2^-$ pentaquark state.  
Hence, $J^P = 1/2^-$ pentaquarks dominated by the $(0s)^5$
configuration  are hardly regarded as a resonance as 
observed in experiments.  

In contrast, for the positive parity $\Thep$ there are 
more configurations available, which together with 
the factor due to the centrifugal barrier suppresses
the transition probability of $\Thep \to KN$.  
The suppression rate depends significantly on the
configuration.  
In the three cases we have studied, one minimizing the spin-flavor 
interaction, spin-color interaction and with the $I=S=0$ 
diquark correlations, the decay widths turned out to be 
70 $\sim$ 10 MeV.  
These values, however, were computed only when color-spin-flavor 
part of the wave functions were properly treated.    
The inclusion of possible spatial correlations and/or with 
more realistic 
configuration mixing would reduce these values.  
Therefore, the width of the pentaquark baryon around 10 MeV
may be explained if the the parity would be positive
when the spin of $\Thep$ is 1/2.  
If it will be of order 1 MeV, then we will still need further
reduction of order 10 for the width, or about 3 for the 
coupling constant.  

Finally, we have pointed out another possibility of 
a narrow resonance of $J^P = 3/2^-$.  
In the hadronic language, this state is dominated by the 
$K^*N$ s-wave bound state.  
Such a picture may not only explain a narrow width of 
the pentaquark state but also provide a simple production 
mechanism as dominated by the $K^*$-exchange.  

As we have seen, the decay width is extremely sensitive 
to the structure of the baryons.  
Therefore, in the quark model more accurate treatment of 
the five-body system 
with reliable model hamiltonian should be desired.  
As several attempts have been reported 
in a recent workshop~\cite{penta}, more development will be 
expecting to appear to clarify the structure of 
the pentaquark state.    

\section*{Acknowledgments}

A.H. thanks H. Toki, T. Nakano, E. Hiyama and M. Kamimura for 
useful discussions.  
His work is supported in part by the Grant for Scientific Research
((C) No.16540252) from the Ministry of Education, Culture, 
Science and Technology, Japan.
The work of M.O. is supported in part by the Grant for 
Scientific Research (B)No.15340072 from the 
Ministry of Education, Culture, Sports, Science and Technology, Japan. 
T. S. is supported by a 21st Century COE Program at Tokyo Tech 
"Nanometer-Scale Quantum 
Physics" 
by the Ministry of Education, Culture, Sports, Science and Technology.

\appendix

\section*{Appendix}

\section{Computation of spectroscopic factor}
\setcounter{equation}{0}
\renewcommand{\theequation}{\Alph{section}-\arabic{equation}}

\subsection{Negative parity $\Theta^+$}

First we start by noting that the color wave function of the $q^4$ 
system must be $[211]_c$ to form the color singlet state with 
$\bar s$ of $[11]_c$.  
This condition should be satisfied not only for the negative 
parity but also for positive parity $\Thep$.  
All four light quarks are then assumed to occupy the lowest s-state 
and therefore the orbital wave function is totally symmetric 
$[4]_o$.  
Therefore, the spin-flavor wave function must have the symmetry 
$[31]_{fs}$ which is combined with the color 
wave function $[211]_c$ to form the totally antisymmetric 
state $[1111]_{csfo}$.  
In the Young diagram, 
\beq
\begin{picture}(3,3) 
\put(0.5,1.5){\framebox(1,1){}}
\put(0.5,0.5){\framebox(1,1){}}
\put(0.5,-0.5){\framebox(1,1){}}
\put(0.5,-1.5){\framebox(1,1){}}
\put(2,-1.5){$_{csfo}$}
\end{picture}
=
\begin{picture}(3,3) 
\put(0.5,1.5){\framebox(1,1){}}
\put(0.5,0.5){\framebox(1,1){}}
\put(0.5,-0.5){\framebox(1,1){}}
\put(0.5,-1.5){\framebox(1,1){}}
\put(2,-1.5){$_{csf}$}
\end{picture}
\cdot 
\begin{picture}(5,3) 
\multiput(0.5,0)(1,0){4}{\framebox(1,1){}}
\put(5,0){$_{o}$}
\end{picture}
\label{yt_csfo}
\eeq

\vspace*{1cm}

\noindent
The subscripts $c, s, f$ and $o$ in the diagram denote color ($c$), 
spin ($s$), flavor ($f$) and orbital ($o$) parts of the wave function. 
Furthermore, center-dot "$\cdot$" denotes the 
inner-product of wave functions in different 
functional space.  
The $csf$ wave function is now decomposed into 
color and spin-flavor part.  
In the Young tableaux with particle number assignment, it 
can be written as
\beq
\begin{picture}(3,3) 
\put(0.5,1){\framebox(1,1){1}}
\put(0.5,0){\framebox(1,1){2}}
\put(0.5,-1){\framebox(1,1){3}}
\put(0.5,-2){\framebox(1,1){4}}
\put(2,-2){$_{csf}$}
\end{picture}
=
\frac{1}{\sqrt{3}}
\left(
\begin{picture}(3,3) 
\put(0.5,0.5){\framebox(1,1){1}}\put(1.5,0.5){\framebox(1,1){4}}
\put(0.5,-0.5){\framebox(1,1){2}}
\put(0.5,-1.5){\framebox(1,1){3}}
\put(2,-1.5){$_c$}
\end{picture}
\cdot
\begin{picture}(4,3) 
\put(0.5, 0){\framebox(1,1){1}}
   \put(1.5,0){\framebox(1,1){2}}
       \put(2.5,0){\framebox(1,1){3}}
\put(0.5,-1){\framebox(1,1){4}}
\put(2,-1){$_{sf}$}
\end{picture}
-
\begin{picture}(3,3) 
\put(0.5,0.5){\framebox(1,1){1}}\put(1.5,0.5){\framebox(1,1){3}}
\put(0.5,-0.5){\framebox(1,1){2}}
\put(0.5,-1.5){\framebox(1,1){4}}
\put(2,-1.5){$_c$}
\end{picture}
\cdot
\begin{picture}(4,3) 
\put(0.5, 0){\framebox(1,1){1}}
   \put(1.5,0){\framebox(1,1){2}}
       \put(2.5,0){\framebox(1,1){4}}
\put(0.5,-1){\framebox(1,1){3}}
\put(2,-1){$_{sf}$}
\end{picture}
+
\begin{picture}(3,3) 
\put(0.5,0.5){\framebox(1,1){1}}\put(1.5,0.5){\framebox(1,1){2}}
\put(0.5,-0.5){\framebox(1,1){3}}
\put(0.5,-1.5){\framebox(1,1){4}}
\put(2,-1.5){$_c$}
\end{picture}
\cdot
\begin{picture}(4,3) 
\put(0.5, 0){\framebox(1,1){1}}
   \put(1.5,0){\framebox(1,1){3}}
       \put(2.5,0){\framebox(1,1){4}}
\put(0.5,-1){\framebox(1,1){2}}
\put(2,-1){$_{sf}$}
\end{picture}
\right)\, .
\label{yd_csf}
\eeq
The Young tableaux is convenient when projecting out such 
a term as containing the quarks 123 
forming the nucleonic component and of 4$\bar 5$ 
kaoninc one.  
The first term is the one of such, where the color 
wave function of 123 is totally antisymmetric $[111]_c$
and spin-flavor part is totally symmetric $[3]_{sf}$.  
Assuming that the $\Thep$ has isospin 0, the flavor 
wave function is expressed by $[22]_f$ and so the only 
possible spin wave function is $[31]_s$.  
Therefore, in the Young tableaux, the spin-flavor 
wave function can be expressed as
\beq
\begin{picture}(4,2)
\put(0.5, 0){\framebox(1,1){1}}
   \put(1.5,0){\framebox(1,1){2}}
       \put(2.5,0){\framebox(1,1){3}}
\put(0.5,-1){\framebox(1,1){4}}
\put(2,-1){$_{sf}$}
\end{picture}
=
\frac{1}{\sqrt{2}}
\left(
\begin{picture}(4,2)
\put(0.5, 0){\framebox(1,1){1}}
   \put(1.5,0){\framebox(1,1){2}}
       \put(2.5,0){\framebox(1,1){4}}
\put(0.5,-1){\framebox(1,1){3}}
\put(2,-1){$_{s}$}
\end{picture}
\cdot
\begin{picture}(3.5,2) 
\put(0.5, 0){\framebox(1,1){1}}
   \put(1.5,0){\framebox(1,1){2}}
\put(0.5,-1){\framebox(1,1){3}}
   \put(1.5,-1){\framebox(1,1){4}}
\put(3,-1){$_{f}$}
\end{picture}
+
\begin{picture}(4,2) 
\put(0.5, 0){\framebox(1,1){1}}
   \put(1.5,0){\framebox(1,1){3}}
       \put(2.5,0){\framebox(1,1){4}}
\put(0.5,-1){\framebox(1,1){2}}
\put(2,-1){$_{s}$}
\end{picture}
\cdot
\begin{picture}(3.5,2)
\put(0.5, 0){\framebox(1,1){1}}
   \put(1.5,0){\framebox(1,1){3}}
\put(0.5,-1){\framebox(1,1){2}}
   \put(1.5,-1){\framebox(1,1){4}}
\put(3,-1){$_{f}$}
\end{picture}
\right) \, . 
\label{yd_sf}
\eeq

Finally the $\bar s$ wave function is multiplied 
to the above $q^4$ wave function.  
The color, spin-flavor wave function of $\bar s$ quark 
is expressed by 
\beq
\bar s \sim 
\left.
\begin{picture}(2.5,2) 
\put(0.5, 0.2){\framebox(1,1){}}
\put(0.5,-0.8){\framebox(1,1){}}
\put(2,-0.8){$_{c}$}
\end{picture}
\cdot 
\begin{picture}(2.5,2) 
\put(0.5, 0.2){\framebox(1,1){}}
\put(0.5,-0.8){\framebox(1,1){}}
\put(2,-0.8){$_{f}$}
\end{picture}
\cdot
\begin{picture}(2.5,1) 
\put(0.5, -0.3){\framebox(1,1){}}
\put(2,-0.3){$_{s}$}
\label{yt_bars}
\end{picture}
\right.
\eeq
This is combined with the $q^4$ wave function to yield the 
$\Thep$ wave function
\beq
|\Thep\ket &=& \frac{1}{\sqrt{3}}
\left(
\begin{picture}(3,3) 
\put(0.5,0.5){\framebox(1,1){1}}\put(1.5,0.5){\framebox(1,1){4}}
\put(0.5,-0.5){\framebox(1,1){2}}
\put(0.5,-1.5){\framebox(1,1){3}}
\put(2,-1.5){$_c$}
\end{picture}
\cdot
\frac{1}{\sqrt{2}}
\left(
\begin{picture}(4,2) 
\put(0.5, 0){\framebox(1,1){1}}
   \put(1.5,0){\framebox(1,1){2}}
       \put(2.5,0){\framebox(1,1){4}}
\put(0.5,-1){\framebox(1,1){3}}
\put(2,-1){$_{s}$}
\end{picture}
\cdot
\begin{picture}(3.5,2) 
\put(0.5, 0){\framebox(1,1){1}}
   \put(1.5,0){\framebox(1,1){2}}
\put(0.5,-1){\framebox(1,1){3}}
   \put(1.5,-1){\framebox(1,1){4}}
\put(3,-1){$_{f}$}
\end{picture}
+
\begin{picture}(4,2) 
\put(0.5, 0){\framebox(1,1){1}}
   \put(1.5,0){\framebox(1,1){3}}
       \put(2.5,0){\framebox(1,1){4}}
\put(0.5,-1){\framebox(1,1){2}}
\put(2,-1){$_{s}$}
\end{picture}
\cdot
\begin{picture}(3.5,2) 
\put(0.5, 0){\framebox(1,1){1}}
   \put(1.5,0){\framebox(1,1){3}}
\put(0.5,-1){\framebox(1,1){2}}
   \put(1.5,-1){\framebox(1,1){4}}
\put(3,-1){$_{f}$}
\end{picture}
\right)
+ \cdots 
\right)
\nonumber \\
& & \hspace*{4cm}
\otimes
\left(
\begin{picture}(2.5,2) 
\put(0.5, 0){\framebox(1,1){}}
\put(0.5,-1){\framebox(1,1){}}
\put(2,-1){$_{c}$}
\end{picture}
\cdot 
\begin{picture}(2.5,2) 
\put(0.5, 0){\framebox(1,1){}}
\put(0.5,-1){\framebox(1,1){}}
\put(2,-1){$_{f}$}
\end{picture}
\cdot
\begin{picture}(2.5,1) 
\put(0.5, -0.5){\framebox(1,1){}}
\put(2,-0.5){$_{s}$}
\end{picture}
\right)
\nonumber \\
&=& 
\frac{1}{\sqrt{3}}
\begin{picture}(3.5,3) 
\put(0.5,0.5){\framebox(1,1){1}}\put(1.5,0.5){\framebox(1,1){4}}
\put(0.5,-0.5){\framebox(1,1){2}}\put(1.5,-0.5){\framebox(1,1){}}
\put(0.5,-1.5){\framebox(1,1){3}}\put(1.5,-1.5){\framebox(1,1){}}
\put(3,-1.5){$_c$}
\end{picture}
\cdot
\frac{1}{\sqrt{2}}
\left(
\begin{picture}(4,2) 
\put(0.5, 0){\framebox(1,1){1}}
   \put(1.5,0){\framebox(1,1){2}}
       \put(2.5,0){\framebox(1,1){4}}
\put(0.5,-1){\framebox(1,1){3}}
   \put(1.5,-1){\framebox(1,1){}}
\put(3,-1){$_{s}$}
\end{picture}
\cdot
\begin{picture}(4.5,2) 
\put(0.5, 0){\framebox(1,1){1}}
   \put(1.5,0){\framebox(1,1){2}}
      \put(2.5,0){\framebox(1,1){}}
\put(0.5,-1){\framebox(1,1){3}}
   \put(1.5,-1){\framebox(1,1){4}}
      \put(2.5,-1){\framebox(1,1){}}
\put(4,-1){$_{f}$}
\end{picture}
+
\begin{picture}(4,2) 
\put(0.5, 0){\framebox(1,1){1}}
   \put(1.5,0){\framebox(1,1){3}}
       \put(2.5,0){\framebox(1,1){4}}
\put(0.5,-1){\framebox(1,1){2}}
   \put(1.5,-1){\framebox(1,1){}}
\put(3,-1){$_{s}$}
\end{picture}
\cdot
\begin{picture}(4.5,2) 
\put(0.5, 0){\framebox(1,1){1}}
   \put(1.5,0){\framebox(1,1){3}}
      \put(2.5,0){\framebox(1,1){}}
\put(0.5,-1){\framebox(1,1){2}}
   \put(1.5,-1){\framebox(1,1){4}}
      \put(2.5,-1){\framebox(1,1){}}
\put(4,-1){$_{f}$}
\end{picture}
\right) \nonumber \\
& & 
\hspace*{2cm} + \cdots \, .  \label{yd_theta_neg}
\eeq
In the first term of this equation, the fourth quark and 
$\bar s$ form the desired color (singlet) and flavor 
(isosinglet) quantum numbers.  
The spin part needs one more step.  
For instance, the spin wave function 
\begin{picture}(4,2) 
\put(0.5, 0){\framebox(1,1){1}}
   \put(1.5,0){\framebox(1,1){2}}
       \put(2.5,0){\framebox(1,1){4}}
\put(0.5,-1){\framebox(1,1){3}}
   \put(1.5,-1){\framebox(1,1){}}
\put(3,-1){$_{s}$}
\end{picture}
has the coupling structure with $S_{1234} = 1$ as\\
\beq
[[S_{123}, S_4]^{S_{1234}}, S_5]^{S_{tot}} 
= [[1/2, 1/2]^1, 1/2]^{1/2}
\label{S1234}
\eeq
which may be recoupled for the kaon with spin 
$S_{45} = 0$:
\beq
[[1/2, 1/2]^1, 1/2]^{1/2} &=&
\sum_J c_J [1/2 [1/2, 1/2]^J ]^{1/2} \, , \nonumber \\
c_0 &=& \frac{\sqrt{3}}{2} \, , \; \; \;  
c_1 = \frac{1}{2} \, .
\label{rcpl_S1234}
\eeq
Here the coefficients $c_0$ and $c_1$ are the amplitude 
for the spin $S_{45} = 0$ and 1 components.  
$S_{45} = 1$ corresponds to the $K^*$ vector meson of 
spin one.  
Therefore, the coupling strength of $K^*$ to the $\Thep$ is 
$1/\sqrt{3}$ of that of $K$ for the negative parity 
$\Thep$.  

Using the results of Eqs.~(\ref{yd_theta_neg}) and 
(\ref{rcpl_S1234}), 
one finds the amplitude of finding the neutron-like $udd$
and kaon($K^+$)-like $u \bar s$ is 
\beq
a = \frac{1}{\sqrt{2}} \frac{1}{\sqrt{3}}\frac{\sqrt{3}}{2}
= \frac{1}{2\sqrt{2}}\, .
\eeq
Note that the first factor $1/\sqrt{2}$ is needed when extracting the 
$K^+n$ component from the isospin zero combination of 
$K^+n$ and $K^0 p$ in the flavor wave function of (\ref{yd_theta_neg}).  
In other words, we have
\beq
|\Theta^+_{csf}\ket 
=
\frac{1}{2\sqrt{2}}|(udd)^n (u\bar s)^{K^+}\ket 
- \frac{1}{2\sqrt{2}}|(uud)^p (d\bar s)^{K^0}\ket 
+ \cdots 
\label{csf_negaiive}
\eeq

\subsection{Positive parity $\Theta^+$}

The wave function for the positive parity $\Theta^+$ 
contains excitation of one unit of orbital angular 
momentum and allows four independent states
with $J^P=1/2^+$ and flavor antidecuplet. 
Assuming that one of the $uudd$ quarks is excited into the 
$l = 1$ p-orbit, the orbital part of the $q^4$ wave function 
takes the symmetry structure $[31]_o$.  
The totally symmetric state $[4]_o$ represents a center-of-mass
motion of the $uudd$ system.  
As in the negative parity case, the spin-flavor-orbital 
wave function has the symmetry $[31]_{sfo}$, and therefore, the 
spin-flavor part can take 
$[4]_{sf}$ or $[22]_{sf}$.  
The spin-flavor decomposition of these states with the flavor 
symmetry $[22]_f$ for antidecuplet is 
\beq
[4]_{sf} &=& [22]_s \cdot [22]_f \nonumber\\
\left[ 22 \right]_{sf} &=& \left(
[22]_s \; \; {\rm or} \; \; [31]_s \; \; {\rm or} \; \; [4]_s 
\right) \cdot [22]_f \, .
\label{sfdecomp_pos}
\eeq
As discussed previously~\cite{Jennings:2003wz}, 
we take the most likely configuration 
for the spin-flavor wave function with the symmetry
$[4]_{sf}$ in which the attraction due to 
the meson-exchange interactions is maximized.  
Such a spin-flavor state may be expressed by the following 
Young tableaux
\beq
\begin{picture}(6,2) 
\put(0.5, -0.5){\framebox(1,1){1}}
   \put(1.5,-0.5){\framebox(1,1){2}}
       \put(2.5,-0.5){\framebox(1,1){3}}
          \put(3.5,-0.5){\framebox(1,1){4}}
\put(5,-0.5){$_{sf}$}
\end{picture}
=
\frac{1}{\sqrt{2}}
\left(
\begin{picture}(3,2) 
\put(0.5, 0){\framebox(1,1){1}}
   \put(1.5,0){\framebox(1,1){2}}
\put(0.5,-1){\framebox(1,1){3}}
   \put(1.5,-1){\framebox(1,1){4}}
\put(3,-1){$_{s}$}
\end{picture}
\cdot
\begin{picture}(4,2) 
\put(0.5, 0){\framebox(1,1){1}}
   \put(1.5,0){\framebox(1,1){2}}
\put(0.5,-1){\framebox(1,1){3}}
   \put(1.5,-1){\framebox(1,1){4}}
\put(3,-1){$_{f}$}
\end{picture}
+
\begin{picture}(3,2) 
\put(0.5, 0){\framebox(1,1){1}}
   \put(1.5,0){\framebox(1,1){3}}
\put(0.5,-1){\framebox(1,1){2}}
   \put(1.5,-1){\framebox(1,1){4}}
\put(3,-1){$_{s}$}
\end{picture}
\cdot
\begin{picture}(4,2) 
\put(0.5, 0){\framebox(1,1){1}}
   \put(1.5,0){\framebox(1,1){3}}
\put(0.5,-1){\framebox(1,1){2}}
   \put(1.5,-1){\framebox(1,1){4}}
\put(3,-1){$_{f}$}
\end{picture}
\right) \, . 
\label{yd_sf_pos}
\eeq
Accordingly, the color-orbital wave function is totally 
antisymmetric:
\beq
\begin{picture}(3,3) 
\put(0.5,1){\framebox(1,1){1}}
\put(0.5,0){\framebox(1,1){2}}
\put(0.5,-1){\framebox(1,1){3}}
\put(0.5,-2){\framebox(1,1){4}}
\put(2,-2){$_{co}$}
\end{picture}
=
\frac{1}{\sqrt{3}}
\left(
\begin{picture}(3,3) 
\put(0.5,0.5){\framebox(1,1){1}}\put(1.5,0.5){\framebox(1,1){4}}
\put(0.5,-0.5){\framebox(1,1){2}}
\put(0.5,-1.5){\framebox(1,1){3}}
\put(2,-1.5){$_c$}
\end{picture}
\cdot
\begin{picture}(4,3) 
\put(0.5, 0){\framebox(1,1){1}}
   \put(1.5,0){\framebox(1,1){2}}
       \put(2.5,0){\framebox(1,1){3}}
\put(0.5,-1){\framebox(1,1){4}}
\put(2,-1){$_{o}$}
\end{picture}
-
\begin{picture}(3,3) 
\put(0.5,0.5){\framebox(1,1){1}}\put(1.5,0.5){\framebox(1,1){3}}
\put(0.5,-0.5){\framebox(1,1){2}}
\put(0.5,-1.5){\framebox(1,1){4}}
\put(2,-1.5){$_c$}
\end{picture}
\cdot
\begin{picture}(4,3) 
\put(0.5, 0){\framebox(1,1){1}}
   \put(1.5,0){\framebox(1,1){2}}
       \put(2.5,0){\framebox(1,1){4}}
\put(0.5,-1){\framebox(1,1){3}}
\put(2,-1){$_{o}$}
\end{picture}
+
\begin{picture}(3,3) 
\put(0.5,0.5){\framebox(1,1){1}}\put(1.5,0.5){\framebox(1,1){2}}
\put(0.5,-0.5){\framebox(1,1){3}}
\put(0.5,-1.5){\framebox(1,1){4}}
\put(2,-1.5){$_c$}
\end{picture}
\cdot
\begin{picture}(4,3) 
\put(0.5, 0){\framebox(1,1){1}}
   \put(1.5,0){\framebox(1,1){3}}
       \put(2.5,0){\framebox(1,1){4}}
\put(0.5,-1){\framebox(1,1){2}}
\put(2,-1){$_{o}$}
\end{picture}
\right)\, .
\label{yd_co_pos}
\eeq

As in the case of negative parity, we need to pick up the term 
where the 123 quarks form a neutron quantum numbers with color singlet and 
totally symmetric in spin-flavor wave function.  
We then need one more decomposition for the orbital part.  
Denoting s ($l = 0$) and p ($l = 1$) states simply by $s$ and $p$, 
the orbital wave function in the relevant term is
\beq
\begin{picture}(4,3) 
\put(0.5, 0){\framebox(1,1){1}}
   \put(1.5,0){\framebox(1,1){2}}
       \put(2.5,0){\framebox(1,1){3}}
\put(0.5,-1){\framebox(1,1){4}}
\put(2,-1){$_{o}$}
\end{picture}
=
\frac{\sqrt{3}}{2} sssp 
- \frac{1}{2} \frac{1}{\sqrt{3}}(pss + sps + ssp)s \, .
\label{dec_sssp}
\eeq
The first term $sssp$ is combined with $\bar s$ state, representing 
a state where the nucleon-like (123) is in the s-state and the kaon 
either moving in p-state or excited intrinsically:
\beq
ps = \frac{1}{\sqrt{2}} \left(
\frac{1}{\sqrt{2}}(ps+sp) + \frac{1}{\sqrt{2}}(ps-sp)
\right) \, .
\label{dec_ps}
\eeq
For the decay 
$\Thep \to K^+n$, we pick up the first term of (\ref{dec_ps}).  
The combination of the first term of (\ref{dec_sssp}) 
and the first term of (\ref{dec_ps}) is referred to as the case (1).  
The second term of (\ref{dec_sssp}) represents that 
the nucleon-like $udd$
is moving in the p-state while the kaon-like $u\bar s$ 
is in the s-state.  
We will refer to this term as the case (2).  
The first and second cases both contribute to $\Thep$ decay 
and are added coherently. 
 
Finally, we need to evaluate the spin rearrangement for $qqqq\bar s$, 
\beq
\begin{picture}(3,2) 
\put(0.5, 0.5){\framebox(1,1){}}\put(1.5,0.5){\framebox(1,1){}}
\put(0.5,-0.5){\framebox(1,1){}}\put(1.5,-0.5){\framebox(1,1){}}
\end{picture}
\otimes
\begin{picture}(2,2) 
\put(0.5, 0){\framebox(1,1){}}
\end{picture}
\, .
\eeq
Using the notation of (\ref{S1234}), it can be done as 
\beq
[[1/2,1/2]^0 , 1/2]^{1/2} 
&=&
\sum_J c^\prime_J [1/2, [1/2, 1/2]^J ]^{1/2}\nonumber \\
c^\prime_0 &=& \frac{1}{2} \, , \; \; \;  
c^\prime_1 = \frac{\sqrt{3}}{2} \, .
\label{c_of_p}
\eeq
Here $c_1^\prime /c_0^\prime = \sqrt{3}$ represents the ratio of the 
$K^*$ coupling to $K$ coupling to the $\Thep$, which is the result
presented by Close and Dudek~\cite{Close:2004tp}.
The probabilities of finding the $KN$ state for 
the cases (1) and (2) in the $\Thep$ wave function are 
\beq
P(1) &=& \left| \frac{1}{\sqrt{3}} \frac{\sqrt{3}}{2}
\frac{1}{\sqrt{2}} \frac{1}{2} \right| ^2 
= \frac{1}{32} \, , \nonumber \\
P(2) &=& \left| \frac{1}{\sqrt{3}} \frac{1}{2}
\frac{1}{2} \right| ^2 
= \frac{1}{48} \, , \nonumber \\
& & P(1) + P(2) = \frac{5}{96} \, .
\label{prb_kn_pos}
\eeq
These results agree with that derived in Ref.~\cite{Carlson:2003xb}.  
Note that these are probabilities of finding 
$KN \sim (K^+n + K^0p)/\sqrt{2}$.  
Therefore, the probabilities of finding $K^+n$ are 
half of them.  
To complete the decomposition of the wave function, we write 
\beq
|\Thep\ket &=& 
\frac{a_1}{\sqrt{2}} | [ [(udd)^n_{l=0, S= 1/2} ]_{j = 1/2},
(\bar s u)^{K^+}_{l=1, S = 0}]_{J=1/2}\ket \nonumber \\
&+& \frac{a_2}{\sqrt{2}} | [ [(udd)^n_{l=1, S= 1/2} ]_{j = 1/2},\ 
(\bar s u)^{K^+}_{l=0, S = 0}]_{J=1/2} \ket
+ ((K^+n) \to (K^0 p)) + \cdots  
\nonumber \\
&=&
\sqrt{\frac{1}{32}} |1\ket - \sqrt{\frac{1}{48}} |2\ket 
+ \cdots
\nonumber \\
&=& 
\sqrt{\frac{5}{96}} 
\left(
\sqrt{\frac{3}{5}} |1\ket - \sqrt{\frac{2}{5}} |2\ket
\right) + \cdots \, .
\label{theta_p1} 
\eeq
Here the states $|1\ket$ and $|2\ket$ are for the states 
of $a_1$ and $a_2$ terms, respectively.  
In these equations, we have shown that the 
$\Theta^+$ contains a components of relative $p$-wave motion 
between the two-quark cluster (meson-like) and three-quark cluster 
(nucleon-like) states.
This reduces to (\ref{theta_p2}).


\end{document}